\documentclass[12pt,preprint]{aastex}
\begin{document}

\title{Can the age discrepancies of neutron stars be circumvented
by an accretion-assisted torque?}

\author{Y. Shi, R. X. Xu}

\affil{School of physics, Peking University, Beijing 100871,
China; rxxu@bac.pku.edu.cn}

%\received{}

\begin{abstract}

It is found that 1E 1207.4-5209 could be a low-mass bare strange
star if its small radius or low altitude cyclotron formation can
be identified.
The age problems of five sources could be solved by a
fossil-disk-assisted torque. The magnetic dipole radiation
dominates the evolution of PSR B1757-24 at present, and the others
are in propeller (or tracking) phases.

\end{abstract}


\section{Introduction}

The age of neutron star is an essential parameter, which is
relevant to the physics of supernova explosion and thereafter the
evolution of stars. However, it is still a big problem now to
determine generally an exact value of age (except the crab
pulsar).
It is a conventional and convenient way to obtain the age for
rotation-powered neutron stars by equalizing the energy lose rate
of spindown to that of magnetodipole radiation, assuming that the
inclination angle between magnetic and rotational axes is
$\alpha=90^{\rm o}$ (e.g., Manchester \& Taylor 1977). The
conclusion keeps quantitatively for any $\alpha$, as long as the
braking torques due to magnetodipole radiation and the unipolar
generator are combined (Xu \& Qiao 2001).
The resultant age, the so-called characteristic age, is $T_{\rm
c}=P/(2{\dot P})$ if the initial period $P_0$ is much smaller than
the present period $P$.
The age, $T_{\rm c}$, is generally considered as the true one for
a neutron star with $P>100$ ms since most newborn neutron stars
could rotate initially at $P_0\sim (20-30)$ ms (e.g., Xu et al.
2002).

It challenges the opinion above that the ages of a few supernova
remnants are inconsistent with $T_{\rm c}$ of their related
isolated stars (Table 1), which implies that some additional
torque mechanisms do contribute to star braking.
Among the five stars, three of them have $T_{\rm c} > 10 T_{\rm
SNR}$ (the age of supernova remnant), and other two $T_{\rm c} \la
2 T_{\rm SNR}$.
In addition, electron cyclotron resonant lines are detected in two
of the five neutron stars (1E 1207.4-5209: Bignami et al. 2003; 1E
2259+568: Iwasawa et al. 1992), but the inferred magnetic fields,
$B_{\rm cyc}$, from which are significantly smaller than that in
the magnetic dipole radiation model, $B_{\rm d}$.
The most prominent one in this age discrepancy issue is 1E
1207.4-5209, which has $T_{\rm c} \ga 30 T_{\rm SNR}$ and $B_{\rm
cyc}\la B_{\rm d}/30$.

One probable and popularly-discussed way to solve the age problem
is through an additional accretion torque (Marsden et al. 2001,
Alpar et al. 2001, Menou et al. 2001).
In this paper, whether an additional accretion torque can possibly
solve the age discrepancy is investigated, including discussions
about possible astrophysical implications.

%\newpage

\section{The case of 1E 1207.4-5209}

The key point in the age discrepancy of 1E 1207.4-5209 is how to
spin down from $P_0\sim 20$ ms to 424 ms in a short time of
$T_{\rm SNR}\sim 7$ kyrs if its true age is $T_{\rm SNR}$.
Certainly the problem disappears if one assumes a long initial
period $P_0\sim 400$ ms or a large braking index $n\sim 50$
(Pavlov et al. 2002); but this is not of Occam's razor since it is
generally believed that rotation-powered radio pulsars born with
$\sim 20$ ms, and brake with index $\la 3$.
May an additional accretion torque help the spindown?
Actually, in an effort to reconcile $B_{\rm d}$ with $B_{\rm
cyc}$, an accretion model for 1E 1207.4-5209 was proposed (Xu et
al. 2003).
However, a very difficulty in the model is how to choose a
time-dependent accretion rate ${\dot M}_{\rm d}(t)$, and to
determine the propeller torque with the rate ${\dot M}_{\rm d}$.

Nevertheless, the propeller phase works in the centrifugal
inhibition regime when $r_{\rm m}>r_{\rm c}$; for a star with mass
$M$ and magnetic moment $\mu$, the corotation radius $r_{\rm
c}=[GM/(4\pi^2)]^{1/3}P^{2/3}$ and the magnetospheric radius $
r_{\rm m}= [\mu^2/({\dot M}_{\rm d}\sqrt{2GM})]^{2/7}$.
To avoid the complex calculations of magnetohydrodynamics, the
rotation energy loss due to propeller torque could be simply
introduced as ${\dot E}_{\rm a}=-G{\dot M}_{\rm d}M/R_{\rm m}$,
based on the energy conservation law.
This is unphysical, but should be an limit for accretion braking.
As $r_{\rm m}$ decreases ($\rightarrow r_{\rm c}^+$), ${\dot
M}_{\rm d}$ increases, and $|{\dot E}_{\rm a}|$ increases too.
Therefore the most efficient spindown (MESD) takes place when
$r_{\rm m}\rightarrow r_{\rm c}^+$.

In a model where the propeller and electromagnetic torques are
combined, in the MESD case, one can derive the period evolution
\begin{equation}
P<1.1 B_{12}^2R_6^4(M/M_\odot)^{-2}(t/{\rm yrs})+P_0~({\rm ms}),
\label{P}
\end{equation}
where $B_{12}=B/(10^{12}$G) and $R_6=R/(10^6$cm). The right hand
of Eq.(\ref{P}) is an upper limit of $P$ because 1, a realistic
accretion rate may not be as high as that of MESD; and 2, the
corresponding braking torque is not so effective.
If 1E 1207.4-5209 is a conventional neutron star with mass $\sim
M_\odot$ and radius $\sim 10^6$ cm, and the line features are
related to cyclotron absorptions near the surface (Xu et al. 2003;
the polar magnetic field is thus $6\times 10^{10}$ G), one has
$P<3.8(t/{\rm kyrs})+P_0$ (ms).

Therefore, assuming 1E 1207.4-5209 has a true age $t\sim 7$ kyrs
and an initial period $P_0\sim 20$ ms, the upper limit of the
present period is $\sim 40$ ms ($\ll P=434$ ms), and the age
discrepancy can then not be solved in the conventional neutron
star model.
However, if 1E 1207.4-5209 is a strange star with low mass, for
instance $R=1$ km (and the mass is thus $\sim 10^{-3}M_\odot$,
since low-mass strange stars have almost a homogenous density
$\sim 4\times 10^{14}$g/cm$^3$; Alcock et al. 1986), the upper
limit is then $P\simeq 110 B_{12}^2(t/{\rm yrs})+P_0$ (ms). In
this case, 1E 1207.4-5209 could spin down to $\sim 2.8$ s during
$\sim 7$ kyrs if its polar magnetic field $6\times 10^{10}$ G.
In fact, the fitted radius of 1E 1207.4-5209 with a blackbody
model is only $\sim 1$km (Mereghetti et al. 1996; Vasisht et al.
1997) although a lager radius is possible if a light-element
atmosphere is applied (Zavlin, Pavlov \& Tr\"umper 1998).
The best-fit tow blackbody model of {\em XMM-Newton} data
indicates an emitting radius $\sim 3$ km for the soft component
with temperature $\sim 200$ eV (Bignami et al. 2003).
Combined with its non-atomic feature spectrum, we may suggest that
1E 1207.4-5209 is a low-mass strange star with bare quark surface
(Xu 2002, Xu et al. 2003).

An alternative possibility is that 1E 1207.4-5209 is a
conventional neutron star, but the cyclotron resonant absorption
forms far away from the surface.
The polar magnetic field is $\sim 6\times 10^{10}[(R+h)/R]^3$G if
the resonant lines form at a height $h$.
From Eq.(\ref{P}), $424 < 1.1\times 7\times 10^3 B_{12}^2+20$, we
estimate a low limit of the polar magnetic field to be $\sim 2.3
\times 10^{11}$ G. This implies that the resonant absorption
region should be at a level of $>16$ km height from the surface.
Certainly, in case of no propeller torque (i.e., the polar
magnetic field is $(1.7-3.6)\times 10^{12}$G), the height of
resonant absorption region is $(30-40)$km.

It is worth noting that 1E 1207.4-5209 could be a low-mass neutron
star with polar magnetic field $B_{12}=0.06$.
From Eq.(\ref{P}) and the conditions of MESD, one has
$R_6^2>4(M/M_\odot)$ for $P_0\sim 20$ ms. This implies a neutron
star with radius $R>10$ km but mass $M<M_\odot$ (e.g., Shapiro \&
Teukolsky 1983).
This result may have difficulties in explaining 1, a non-atomic
spectrum (Xu et al. 2003), and 2, a possible small radius observed
(Mereghetti et al. 1996; Vasisht et al. 1997; Bignami et al. 2003)
and even the fitting result of a neutron star with 10 km and
$1.4M_\odot$ (Zavlin et al. 1998).

\section{Other sources}

If other sources list in Table 1 are neutron stars with $R_6=1$
and $M=M_\odot$, the low limits of polar magnetic fields are
$6.1\times 10^{11}$ G for 1E 2259+586,
$4.9\times 10^{10}$ G for PSR B1757-24,
$1.6\times 10^{11}$ G for PSR J1811-1925,
and $(2.5-5.6)\times 10^{11}$ G for PSR J1846-0258.
Among these sources, only possible cyclotron absorption is found
in 1E 2259+586, and the limit field is within the range of that
inferred from cyclotron line. This suggests that the cyclotron
resonant may take place just above the stellar surface.

An interesting question is: how much mass could accrete during the
propeller phase in case of MESD?
Certainly, only a very small part of this matter can accrete onto
the stellar surface.
When MESD works, one obtains the accretion rate
${\dot M}_{\rm d}\sim 2^{11/6}\pi^{7/3}\mu^2(GM)^{-5/3}P^{-7/3}$
from $r_{\rm m}\simeq r_{\rm c}$. If the quantities are re-scaled,
$m=M_{\rm d}/M_\odot$, $\tau=t/{\rm yr}$, and $p=P/{\rm ms}$, one
has
${\rm d}m/{\rm d}\tau\sim 0.24~
\mu_{30}^2(M/M_\odot)^{-5/3}p^{-7/3}$.
Combining with Eq.(\ref{P}), one comes to
\begin{equation}
m<\int_0^\infty {\rm d}\tau =
0.16(M/M_\odot)^{-2/3}I_{45}p_0^{-4/3},
\label{m}
\end{equation}
where $p_0=P_0/{\rm ms}$. Note that the upper limit of accretion
mass, $M_{\rm d}$, in the right hand of Eq.(\ref{m}) does not
depend on the magnetic momentum. Typically for $p_0=20$, the upper
limit of accretion mass is $2.95\times 10^{-3}M_\odot$, which is
reasonable since the amount of the fall-back material after
supernova explosion could be as high as $0.1M_\odot$ (Lin, Woosley
\& Bodenheimer 1991; Chevalier 1989).
Due to $r$-mode instability, a nascent neutron star may loss
rapidly its angular momentum though gravitational radiation if the
initial period is less than $\sim 3-5$ ms (Andersson \& Kokkotas
2001).
The upper limit of accretion mass in case of MESD is $\sim
0.04M_\odot$ for $p_0=3$.
These results indicate that the fall-back matter is enough to
brake the center stars by propeller torque in MESD case.

\section{A model with self-similar accretion rate}

Though the study about MESD torque provides some useful
information on the accretion model, including the appropriate
magnetic field and the mass of fall-back disk around a neutron
star, the accretion rate of MESD torque is questionable in
realistic cases.
After a dynamical time a fossil disk may form. For a
viscosity-driven disk, the accretion could be in a self-similar
way, with an accretion rate of (Cannizzo, Lee \& Goodman 1990)
\begin{equation}
{\dot m}={\dot m_0},~0<t<T;~~
{\dot m}={\dot m_0}(t/T)^{-\alpha},~t\geq T,
\label{dotm}
\end{equation}
where $T$ is of order the dynamical time, ${\dot m}={\rm d}m/{\rm
d}t$.
Assuming $T\sim 1$ms, an initial disk mass $\sim 0.006M_\odot$,
and an opacity dominated by electron scattering ($\alpha=7/6$),
Chatterjee, Hernquist and Narayan (2000) develop a first detailed
model of fossil disk accretion for anomalous X-ray pulsars (AXPs).
However it is noted by Francischelli and Wijers (2002) that
Kramers opacity may prevail in the fossil disk (i.e.,
$\alpha=1.25$).
In the regime of conventional neutron stars, we will calculate the
accretion torque through the realistic accretion rate of
Eq.(\ref{dotm}), assuming $\alpha=1.25$ and $T=1$ms, with the
inclusion of magnetic dipole radiation. The spinup/spindown torque
proposed by Menou et al. (1999),
\begin{equation}
{\dot J}=2{\dot M}_{\rm d}r_{\rm m}^2\Omega_{\rm k}(r_{\rm
m})[1-\Omega/\Omega_{\rm k}(r_{\rm m})],
\end{equation}
is applied for the action of fossil disk in the model, where
$\Omega=2\pi/P$, $\Omega_{\rm k}(r_{\rm m})$ is the Keplerian
angular velocity at the magnetospheric boundary.

One may compute the accretion rate, $\dot m$, as well as the spin
evolution $P(t)$. The total disk mass could be
$m=\int_0^\infty{\dot m}{\rm d}t$. It is worth noting that the
disk mass obtained in this way could be much larger than that in
MESD case because of the inclusion of the high accretion in the
initial period.
%
%However this initial accretion is insensitive to the long
%evolution history at late time, since the spinup is not very
%effective due to the angular momentum loss via gravitational
%radiation (e.g., Andersson \& Kokkotas 2001).

We think that the accretion rate characterized by Eq.(\ref{dotm})
is on average in a sense. The period derivative, $\dot P$, may be
affected by dynamical instabilities or some stochastic processes,
whereas the period, $P$, is of the integration over a very long
time.
We therefore calculate $P(T_{\rm SNR})$ for any disk mass, $m$,
and polar magnetic field, $B_{12}$, of neutron stars.
For 1E 1207.4-5209, the calculated contour of period relative
error, $|P(T_{\rm SNR})-P|/P$ ($P=0.424$s for 1E 1207.4-5209), is
shown in Fig.1.
A reasonable parameter set (disk mass $m$ and polar magnetic field
$B_{12}$) is chosen if the following criteria are met: 1, the
period relative error is smaller; 2, $m<0.1$; 3, $B_{\rm
12,cyc}<B_{12}<B_{\rm 12,d}$ (Table 1). We then have $m=0.054$ and
$B_{12}=3.55$.
The parameter sets for other sources can also be obtained in this
way, which are listed on Fig.2 (except for PSR J1811-1925).
The period evolution curves, with these parameters, are drawn in
Fig.2. Note these curves do not change significantly if the
parameter sets shift reasonably.

The heights, $h$, of cyclotron resonant scattering regions can be
obtained, based on the differences of parametric magnetic field,
$B_{12}$, and the field inferred from absorption features, $B_{\rm
12,cyc}$.
It is found that $h\sim 29$ km and $\sim (8.8-15)$ km for 1E
1207.4-5209 and 1E 2259+586, respectively, in the model.

Whether or not the disk will influence the spindown of the neutron
star or suppress the radio emission will depend on the location of
$r_{\rm m}$ relative to the light cylinder radius, $r_{\rm L}$,
and the corotation radius $r_{\rm c}$ (Chatterjee et al. 2000).
Magnetic dipole radiation dominates, and the disk and the star
will effectively evolve independently if $r_{\rm m}>r_{\rm L}$;
but in other cases, accretion onto the star will lead to
accretion-induced X-ray emission with radio quiet.
We see from Fig.2 that the condition of $r_{\rm m}>r_{\rm L}$ is
satisfied only for PSR B1757-24 when it is older than $\sim 10^3$
years. We are therefore not surprise that PSR B1757-24 is now
radio loud whereas the others (1E 1207.4-5209, 1E 2259+586, and
PSR J1846-0258) are radio quiet.
The AXP 1E 2259+586 is in a tracking phase, and we expect other
two (1E 1207.4-5209 and PSR J1846-0258) will evolve to be AXPs
when they are in tracking phases too.

PSR J1811-1925 is an interesting exception among the five sources,
whose age is certain if it has physical association with the
remnant of a supernova recorded in A.D. 386. In its calculated
contour, we can only choose \{$B_{12}=2.6, m=0.1$\} or
\{$B_{12}=1.6, m=2.2$\}; both the parameter sets are not
reasonable (i.e., can not meet those 3 criteria).
This may imply that the accretion of PSR J1811-1925 is not
self-similar. Recalling that the low limit of polar field is only
$1.6\times 10^{11}$G if in MESD case, we could suggest that PSR
J1811-1925 has a field within $(1.6-17)\times 10^{11}$G, with an
accretion stronger than that of Eq.(\ref{dotm}) but weaker than
that in MESD case. This result hints that PSR J1811-1925 is radio
quiet (Crawford et al. 1998).
In addition, the parametric field, $B_{12}=3.55$, chosen for 1E
1207.4-5209, which is close to $B_{\rm 12,d}=3.6$, would also
indicate that the real accretion is not described by
Eq.(\ref{dotm}).
In fact the accretion of Eq.(\ref{dotm}) is for the capture of
material by black holes where magnetic field is not important,
which could differ from that for neutron stars with strong fields.

\section{Conclusions and Discussions}

The possibility of solving the age discrepancy by an
accretion-assisted torque is discussed. We find that:
1, 1E 1207.4-5209 can not be a neutron star, but a low-mass bare
strange star, if the cyclotron resonant region is near the polar
cap with a magnetic field of $6\times 10^{10}$G; whereas it could
be a conventional neutron star if the cyclotron lines form at a
height of $(16-40)$km. An identification of smaller radius or low
altitude cyclotron formation favors a low-mass bare strange star
model for 1E 1207.4-5209.
2, Among the five sources with age problems, the magnetic dipole
radiation dominates the evolution of PSR B1757-24 at present, and
the others are in propeller (or tracking) phases.
3, The real accretion around these sources may differ from a
self-similar one (Eq.(\ref{dotm})), at least for PSR J1811-1925.
4, By a calculation with self-similar accretion, it is suggested
that PSR J1846-0258 and 1E 1207.4-5209 (and probably PSR
J1811-1925) would evolve to be anomalous X-ray pulsars in the
future.

The debris disks formed following supernovae are currently
conjectured also for interpreting other astrophysical phenomena,
e.g, anomalous X-ray pulsars and soft $\gamma-$ray repeaters.
Factually, these disks around the sources could be bright in a
wide spectral range.
Recent discoveries of possible optical and near-infrared emission
from a few AXPs may be a hint of such kind fallback accretion
disks (1E 2259+586: Hulleman et al. 2001; 1RXS J170849-400910:
Israel et al. 2002; 1E 1048.1-5937: Wang \& Chakrabarty 2002).
Although a comparison of optical and near-infrared observations
with theoretical predictions of spectra of disks around neutron
stars (Perna et al. 2000) have helped rule out the presence of
disks in some cases, more detailed studies in this aspect is still
necessary, which may be an effective way to test the fossil-disk
model for young neutron stars.

\vspace{0.2cm} \noindent {\it Acknowledgments}:
The helpful suggestions and discussions from an anonymous referee
are sincerely acknowledged.
This work is supported by National Nature Sciences Foundation of
China (10273001) and the Special Funds for Major State Basic
Research Projects of China (G2000077602).

\clearpage

%%%%%%%%%%%%%%%%%%%%%%%%% Table 1 %%%%%%%%%%%%%%%%%%%%%%%%
%
\begin{deluxetable}{llccccccl}
\tabletypesize{\scriptsize} \tablewidth{0pt} \tablecaption{List of
neutron stars with age discrepancies} \tablehead{ \colhead{Stars}
& \colhead{SNRs} & \colhead{$P$(s)} &
\colhead{$\dot{P}$(10$^{-13}$s/s)}  & \colhead{$T_{\rm c}(10^3{\rm
y})$} & \colhead{$T_{\rm SNR}(10^3{\rm y})$} & $B_{\rm 12,d}^*$ &
$B_{\rm 12,cyc}^{\S}$ & \colhead{Ref.} } \startdata
1E 1207.4-5209 & PKS 1209-51/52 & 0.424 & 0.07-0.3 & 200-900 &
 $\sim 7$ & 1.7-3.6 & 0.06 & 1,2\\
1E 2259+586$^{\dag}$ & CTB 109 & 6.98 & 4.84 & 228 & 17
 & 59 & 0.4-0.9 & 3,4,5\\
PSR B1757-24 & G5.4-1.2 & 0.125 & 1.28 & 16 & $>39$
 & 4.0 & --- & 6\\
PSR J1811-1925 & G11.2-0.3 & 0.065 & 0.44 & 24 & 1.6
 & 1.7 & --- & 7\\
PSR J1846-0258 & Kes 75 & 0.325 & 71 & 0.72 & 0.9-4.3
 & 49 & --- & 8\\
\enddata
\tablenotetext{*}{The magnetic fields in the magnetic dipole
radiation model: $B_{\rm d}=3.2\times 10^{19}\sqrt{P {\dot P}}$ G,
$B_{\rm 12,d}=B/(10^{12}{\rm G})$.}
\tablenotetext{\S}{The magnetic fields (in unit of $10^{12}$ G)
derived from spectral features as cyclotron resonant scattering.
No gravitation redshift is included here.}
\tablenotetext{\dag}{An anomalous X-ray pulsar.}
\tablerefs{%
1, Pavlov et al. (2002); 2, Bignami et al. (2002); 3, Gavriil \&
Kaspi (2002); 4, Hughes et al. (1981); 5, Iwasawa et al. (1992);
6, Marsden et al. (2001); 7, Torii et al. (1999); 8, Gotthelf et
al. (2000).
}
\end{deluxetable}
%%%%%%%%%%%%%%%%%%%%%%%%%%%%%%%%%%%%%%%%%%%%%%%%%%%%%%%%%

\clearpage

%%%%%%%%%%%%%%%%%%%%%%%%% Fig.1 %%%%%%%%%%%%%%%%%%%%%%%%%
%
\begin{figure}
\plotone{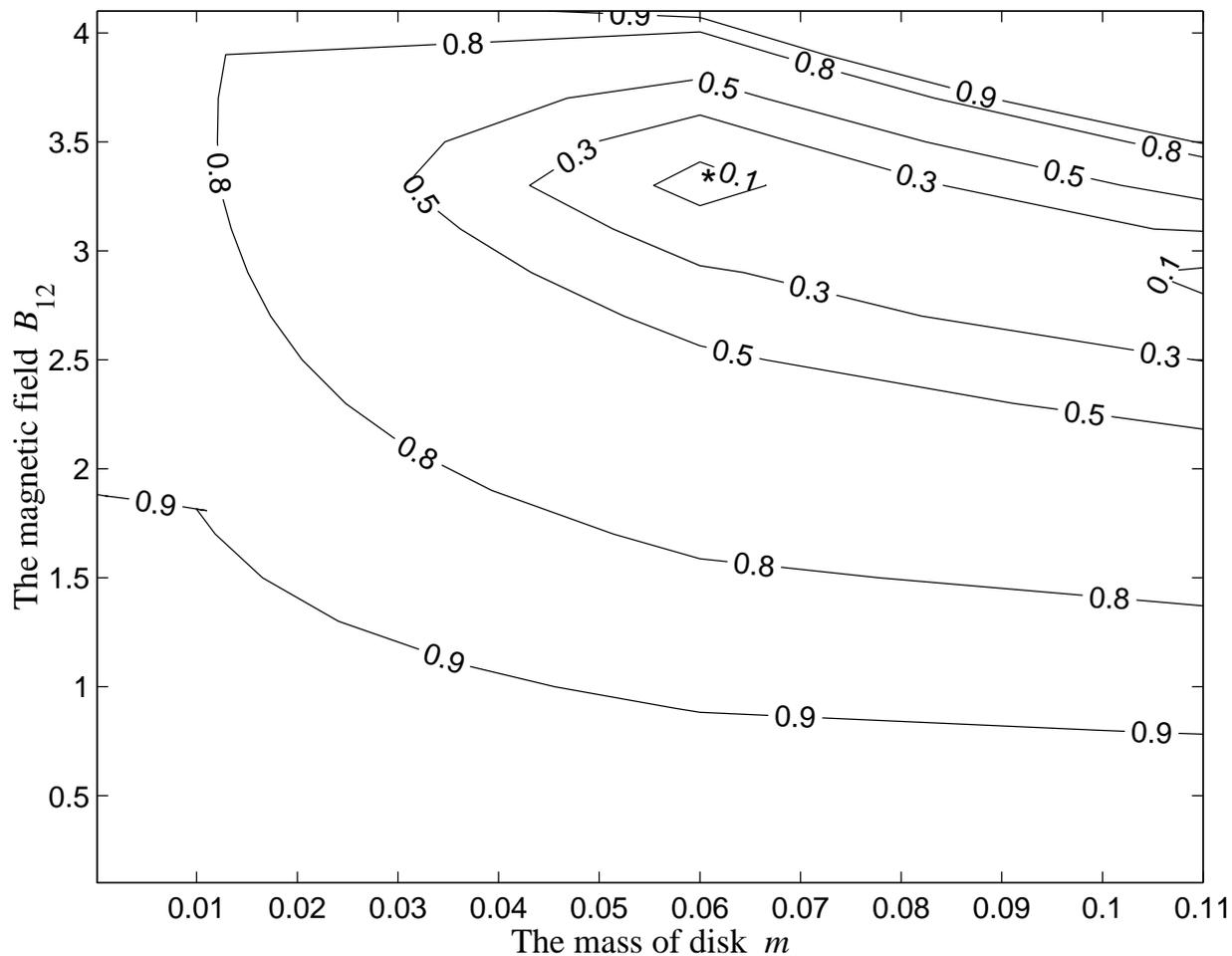} \caption{The contour of the relative error of
period for 1E1207.4-5209 in a model with self-similar accretion.
The star sign indicates the parametric position we choose for a
reasonable neutron star and the fossil disk around it. The disk
mass $m$ is in $M_\odot$ and the polar magnetic field $B_{12}$ in
$10^{12}$G.
\label{fig1}}
\end{figure}
%%%%%%%%%%%%%%%%%%%%%%%%%%%%%%%%%%%%%%%%%%%%%%%%%%%%%%%%%

%%%%%%%%%%%%%%%%%%%%%%%%% Fig.2 %%%%%%%%%%%%%%%%%%%%%%%%%
%
\begin{figure}
\plotone{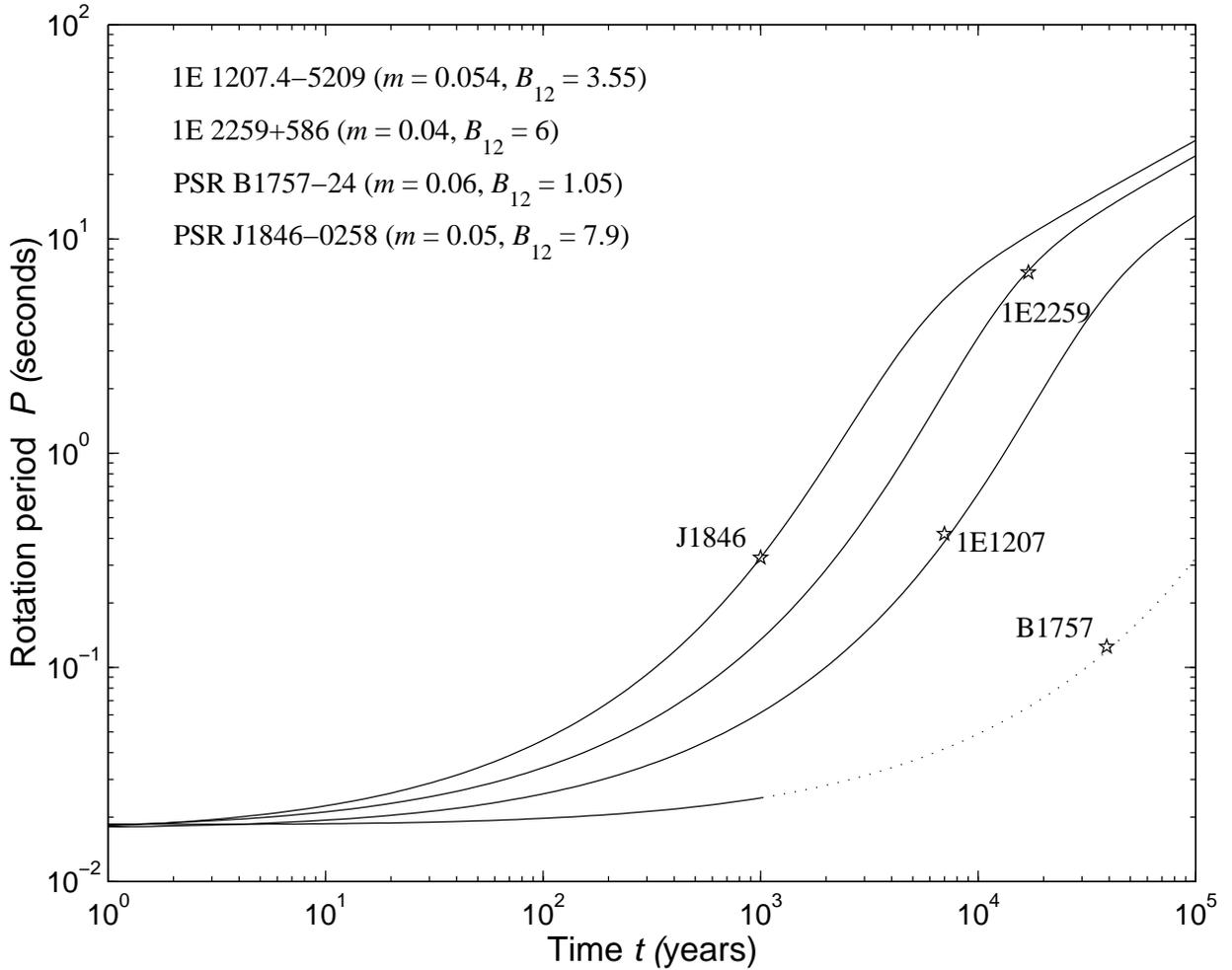} \caption{The period evolution in a model with
self-similar accretion for the four pulsars labelled. The
parameter sets used to calculate the curves are also listed. The
solid part of the curves indicates that the accretion torque works
(i.e., the magnetospheric radius is smaller than that of light
cylinder, $r_{\rm m}<r_{\rm L}$), while the dashed part of PSR
B1757-24 means that the pulsar and the fossil disk could evolve
independently (i.e., $r_{\rm m}>r_{\rm L}$).
\label{fig2}}
\end{figure}
%%%%%%%%%%%%%%%%%%%%%%%%%%%%%%%%%%%%%%%%%%%%%%%%%%%%%%%%%

\end{document}